\documentclass[10pt]{article}

\usepackage{amssymb}
\def\lddots{\mathinner{\mkern1mu\raise1pt\hbox{.}\mkern2mu
\raise4pt\hbox{.}\mkern2mu\raise7pt\vbox{\kern7pt\hbox{.}}\mkern1mu}} 
\makeatletter
\def\numberbysection{\@addtoreset{equation}{section}
\def\theequation{\thesection.\arabic{equation}}}
\makeatother

\numberbysection

\newcommand{\be}{\begin{eqnarray}}
\newcommand{\ee}{\end{eqnarray}}
\newcommand{\non}{\nonumber}

\begin{document}

%\begin{titlepage}
\vskip 0.4cm
\strut\hfill{LAPTH-conf-1115/05}
\vskip 0.8cm
\begin{center}

%\begin{center}

{\large \bf{On the symmetries of integrable systems with boundaries}}
\vspace{7mm}

{ANASTASIA DOIKOU\footnote{e-mail: doikou@lapp.in2p3.fr}}

\vspace{3mm}

{\large \emph{LAPTH, Annecy-Le-Vieux, B.P. 110, F-74941, France}}

\end{center}

%\vfill

\vspace{8mm}

\begin{abstract}
We employ appropriate realizations of the affine Hecke algebra and we recover previously known non-diagonal solutions of the reflection equation for the $U_{q}(\widehat{gl_n})$ case. With the help of linear intertwining relations involving the aforementioned solutions of the reflection equation, the symmetry of the open spin chain with a particular choice of the left boundary is exhibited. The symmetry of the corresponding local Hamiltonian is also explored.
\end{abstract}

%\vfill
%MSC number: 81R50, 17B37
%\rightline{LAPTH-xxxx/05}
%\rightline{hep-th/04}
%\rightline{August 2005}
%\baselineskip=16pt
%\end{titlepage}

\section{Introduction}

In order to construct an open spin chain one needs two basic building blocks, namely the $R$ matrix acting on $V^{ \otimes 2}$, satisfying the Yang-Baxter equation \cite{baxter} \be
R_{12}(\lambda_{1}-\lambda_{2})\ R_{13}(\lambda_{1})\
R_{23}(\lambda_{2}) =R_{23}(\lambda_{2})\ R_{13}(\lambda_{1})\ R_{12}(\lambda_{1}-\lambda_{2}), \label{YBE} \ee 
and the $K$ matrix acting on $V$, and obeying the reflection equation \cite{cherednik}
\begin{equation} R_{12}(\lambda_{1} -\lambda_{2})\ 
K_{1}(\lambda_{1})\  R_{21}(\lambda_{1}+\lambda_{2})\ K_{2}(\lambda_{2})  = 
 K_{2}(\lambda_{2})\ R_{12}(\lambda_{1}+\lambda_{2})\ K_{1}(\lambda_{1})\ R_{21}(\lambda_{1}-\lambda_{2}). \label{re} \end{equation} 
We shall consider henceforth a particular solution of the Yang--Baxter equation, that is the $R$ matrix associated to $U_{q}(\widehat{gl_{n}})$ \cite{jimbo}
\be R(\lambda) = a(\lambda) \sum_{i=1}^{n} \hat e_{ii} \otimes \hat e_{ii}+ b(\lambda) \sum_{i\neq j=1}^{n} \hat e_{ii} \otimes \hat e_{jj} + c \sum_{i\neq j =1}^{n} e^{ -sgn(i-j)\lambda} \hat e_{ij}\otimes \hat e_{ji}\label{r} \ee where \be a(\lambda)=\sinh (\lambda+i\mu), ~~~b(\lambda) = \sinh \lambda, ~~~c = \sinh i \mu. \label{rhp} \ee

\section{Affine Hecke algebra and solutions of the reflection equation} 

Given the structural similarity between the defining relations of the affine Hecke algebra, and the Yang--Baxter and reflection equations we shall employ representations of the affine Hecke algebra in order to derive solutions of the reflection equation \cite{levymartin, doikoumartin, doikou2}. We shall henceforth restrict our attention to quotients of the affine Hecke algebra \cite{levymartin, doikoumartin, doikou2}.\\
\\
{\bf Definition 1.} {\it  A quotient of the affine Hecke algebra, called the $B$-type Hecke algebra ${\cal B}_{N}(q,Q)$, is defined by generators ${\cal U}_{l}$, $~l\in\{1, \ldots N-1\}$, $~{\cal U}_{0}$ satisfying: \be  {\cal U}_{l}\ {\cal U}_{l} = \delta\  {\cal U}_{l},
~~~{\cal U}_{l}\ {\cal U}_{l+1}\ {\cal U}_{l} - {\cal U}_{l}= {\cal U}_{l+1}\ {\cal U}_{l}\ {\cal U}_{l+1}- {\cal U}_{l+1} 
\label{hecke} \ee \be {\cal U}_{0}\ {\cal U}_{0} = \delta_{0}\ {\cal U}_{0}, ~~~
\Big [{\cal U}_{j},\ {\cal U}_{l} \Big ] =0, ~~~|j-l|>1, \label{cons} \ee \be {\cal U}_{1}\ {\cal U}_{0}\ {\cal U}_{1}\ {\cal U}_{0} - \kappa\ {\cal U}_{1}\ {\cal U}_{0} = {\cal U}_{0}\ {\cal U}_{1}\ {\cal U}_{0}\ {\cal U}_{1}\ -\kappa\ {\cal U}_{0}\ {\cal U}_{1}  \label{hecke0} \ee  $\delta= -q-q^{-1}$ ($q=e^{i\mu}$), $~\delta_{0} =-(Q+Q^{-1})$ and $\kappa = q Q^{-1} +q^{-1} Q$.} We are free to renormalize ${\cal U}_{0}$ and consequently $\delta_{0}$ and $\kappa$. In fact, if we relax the first of the relations (\ref{cons}) we recover the affine Hecke algebra. Moreover, the algebra defined by the generators ${\cal U}_{l}$ ($l \neq 0$) is the Hecke algebra ${\cal H}_{N}(q)$.

It was shown in \cite{levymartin, doikoumartin} that tensor representations of quotients of the affine Hecke algebra provide solutions to the reflection equation. For our purposes here we shall make use of the following \cite{doikou2}:\\
\\
{\bf Proposition 1.} {\it Tensor representations of $H_{N}(q)$ that extend to $B_{N}(q,Q)$, $\rho: {\cal B}_{N}(q, Q) \to \mbox{End}(V^{\otimes N})$ provide solutions to the reflection equation, i.e. \be K (\lambda) = x(\lambda)I +y(\lambda) \rho({\cal U}_{0}), \label{ansatz} \ee with \be x(\lambda)= -\delta_{0}\cosh (2\lambda +i \mu) -
\kappa \cosh 2 \lambda -\cosh 2 i\mu \zeta,~~~~
y(\lambda)=2 \sinh 2\lambda\  \sinh i\mu. \label{ansatz2} \ee} {\it Proof}: The proof is straightforward. The values of $x(\lambda)$ and $y(\lambda)$ can be found by direct computation, by substituting the ansatz (\ref{ansatz}) in (\ref{re}) and also using equations (\ref{hecke})--(\ref{hecke0})\\
\\
The representation of the Hecke algebra that provides the $R$ matrix (\ref{r}) is given by \cite{jimbo}, 
$\rho: {\cal H}_{N}(q) \to \mbox{End}((C^{n})^{\otimes N})$ such that
\be \rho({\cal U}_{l}) = I \otimes \ldots \otimes U \otimes \ \ldots \otimes  I \label{sol3} \ee acting non-trivially on $ V_{l} \otimes  V_{l+1}$, with \be U = \sum_{i \neq j =1}^{n}(\hat e_{ij} \otimes \hat e_{ji} - q^{-sgn(i-j)} \hat e_{ii} \otimes \hat e_{jj}). \label{sol} \ee  
The above representation may be extended to the $B$-type Hecke algebra \cite{doikou2} $\rho: {\cal B}_{N}(q,\ Q) \to \mbox{End}((C^{n})^{\otimes N})$  with (\ref{sol3}), (\ref{sol}) and \be  \rho({\cal U}_{0}) = {1 \over 2 i \sinh i\mu} U_{0} \otimes  I \ldots \otimes I \label{solb1} \ee acting non-trivially on $V_{1}$ and \be U_{0}= -Q^{-1} \hat e_{11} -Q \hat e_{nn} +\hat e_{1n}+ \hat e_{n1}. \label{solb} \ee It is convenient to set $Q=i e^{i \mu m}$, then by substituting (\ref{solb}) in the ansatz (\ref{ansatz}) we obtain the $n \times n$ $K$ matrix:
\be &&K_{11}(\lambda) = e^{2 \lambda} \cosh i\mu m - \cosh 2 i \mu \zeta, ~~~K_{nn}(\lambda)=e^{-2 \lambda} \cosh i\mu m - \cosh 2i \mu \zeta \non\\ && K_{1n}(\lambda)=K_{n1}(\lambda)=-i\sinh 2 \lambda,\non\\ && K_{jj}(\lambda) = \cosh (2\lambda +i m \mu) - \cosh 2i \mu \zeta, ~~ j \in \{2, \ldots ,n-1 \}.\label{k}\ee The later matrix coincides with the one found in '95 by Abad and Rios, subject to parameter identifications \cite{doikou2}.
 
\section{The reflection algebra and the open spin chain}     

Having at our disposal c-number solutions of (\ref{re}) we may build the more general form of solution of (\ref{re}) as argued in \cite{sklyanin}. To achieve this it is necessary to define the following objects:
\be
{\cal L}(\lambda)= e^{\lambda} {\cal L}^{+} - e^{-\lambda} {\cal
L}^{-}, \label{lh0} \ee with the matrices ${\cal L}^{+}$ (${\cal L}^{-}$) being upper (lower) triangular, and ${\cal L}$ satisfies: 
\be  R_{ab}(\lambda_{1} -\lambda_{2})\ {\cal L}_{a}(\lambda_{1})\  {\cal L}_{b}(\lambda_{2}) =
{\cal L}_{b}(\lambda_{2})\  {\cal L}_{a}(\lambda_{1})\  R_{ab}(\lambda_{1} -\lambda_{2}). \label{funda} \ee
Define also 
\be \hat {\cal L}(\lambda) = {\cal L}^{-1}(-\lambda). \ee The more general solution of (\ref{re}) is then  given by \cite{sklyanin}:
\be {\cal K}(\lambda) = {\cal L}(\lambda-\Theta)\  (K(\lambda)\otimes  I)\  \hat {\cal L}(\lambda +\Theta), \label{gensol} \ee where $K$ is the c-number solution of the reflection equation, $\Theta$ is a constant and hereafter will be considered zero. The entries of ${\cal K}$ are elements of the so called reflection algebra ${\cal R}$, with exchange relations dictated by the algebraic constraints (\ref{re}), (see also \cite{sklyanin}). 

One may easily show that all the elements of the reflection algebra `commute' with the solutions of the reflection equation (see also \cite{dema}). Let $\pi_{\lambda}$ be the evaluation representation  $\pi_{\lambda}: U_{q}(\widehat{gl_{n}}) \to 
\mbox{End}(C^{n})$ \cite{jimbo}. Then by acting with the evaluation representation on the second space of (\ref{gensol}) it follows \be \pi_{\lambda} ({\cal K}_{ij}(\lambda'))\ K(\lambda) =  K(\lambda)\ \pi_{-\lambda} ({\cal K}_{ij}(\lambda')), ~~~i,\ j \in \{1, \ldots,n \}. \label{bcomm}\ee The reflection algebra is also endowed with a coproduct inherited essentially from ${\cal A}$ \cite{dema, doikou2}, i.e. $\Delta: {\cal R} \to{\cal R} \otimes {\cal A}$, such that \be \Delta( {\cal K}_{ij}(\lambda)) = \sum_{k,l=1}^n {\cal K}_{kl}(\lambda) \otimes {\cal L}_{ik}(\lambda)\ \hat {\cal L}_{lj}(\lambda)~~~i,~j \in \{1, \ldots, n\}. \label{coc} \ee  The open spin chain may be constructed following the generalized QISM \cite{sklyanin}. We first need to define \be T_{0}(\lambda) =   {\cal L}_{0N}(\lambda)\ldots   {\cal L}_{01}(\lambda),~~~~~~ \hat T_{0}(\lambda) =  \hat {\cal L}_{01}(\lambda)\ldots  \hat {\cal L}_{0N}(\lambda) \label{th} \ee usually the space denoted by 
0 is called auxiliary, whereas the spaces denoted by $1, \ldots, N$ are called quantum. The general tensor type solution of the (\ref{re}) takes the form \be {\cal T}_{0}(\lambda) = T_{0}(\lambda)\ K_{0}^{(r)}(\lambda)\ \hat T_{0}(\lambda). \label{transfer0} \ee Notice that relations similar to (\ref{bcomm}) may be derived for ${\cal T}$ \cite{doikou2, doikou1}, i.e. \be (\pi_{\lambda} \otimes
\mbox{id}^{\otimes N})\Delta^{'(N+1)}({\cal K}_{ij}(\lambda'))\ {\cal T}(\lambda) = {\cal T}(\lambda)\ (\pi_{-\lambda} \otimes
\mbox{id}^{\otimes N})\Delta^{'(N+1)}( {\cal K}_{ij}(\lambda')), \label{it} \ee where \be \Delta' = \Pi \circ \Delta, ~~~~\Pi:~a\otimes b \to b\otimes a. \ee We can now introduce the transfer matrix of the open spin chain \cite{sklyanin}, which may be written as \be t(\lambda) = Tr_{0}\   \Big \{ M_{0}\ K_{0}^{(l)}(\lambda)\ {\cal T}_{0}(\lambda) \Big \}.  \label{transfer00} \ee where $M_{ij}= e^{i\mu(n-2j+1)}\delta_{ij}$, $~K^{(r)}$ is a solution (\ref{k}) of the reflection equation (\ref{re}), and here we consider $~K^{(l)}(\lambda) = I$.

It can be proved using the fact that ${\cal T}$ is a solution of the reflection equation (\ref{re}) that \cite{sklyanin} \be \Big [ t(\lambda),\ t(\lambda') \Big ]=0, \ee  which ensures that the open spin chain (\ref{transfer00}) is integrable.\\
\\
{\bf The Hamiltonian:} It is useful to write down the Hamiltonian of the open spin chain. For this purpose we should restrict our attention in the case where the evaluation representation $\pi_{0}$ acts on the quantum spaces of the spin chain well, then $~{\cal L}(\lambda) \to R(\lambda)$, $~\hat {\cal L}(\lambda) \to \hat R (\lambda) = R^t(\lambda)~$ ($t$ denotes total transposition).  The Hamiltonian is given by
\be {\cal H} = -{(\sinh i\mu)^{-2N+1} \over {4 x(0)}} \Big( tr_{0} M_{0} \Big )^{-1}\ 
\ {d \over d \lambda} t(\lambda)\vert_{\lambda =0} \label{H0} \ee  and it may be written exclusively in terms of the generators of ${\cal B}_{N}$ as
\be  {\cal H} = -\frac{1}{2}
\sum_{l=1}^{N-1} \rho({\cal U}_{l}) 
- \frac{\sinh i\mu\ y'(0)}{4  x(0)} \rho({\cal U}_{0})+ c \rho(1). \label{ht} \ee

\section{The boundary symmetry}  

From  the asymptotic behaviour of ${\cal T}$ we obtain the `boundary non-local' charges (entries of ${\cal T}(\lambda \to \infty)$) \cite{doikou2, doikou1}\be  &&{\cal T}_{11}^{+(N)} =2 \cosh i\mu m\ T_{11}^{+(N)}\ \hat T_{11}^{+(N)} -i T_{1 n}^{+(N)}\ \hat T_{11}^{+(N)} -i T_{11}^{+(N)}\ \hat  T_{n1}^{+(N)}+e^{i\mu m}  \sum_{j=2}^{n-1}  T_{1j}^{+(N)}\ \hat T_{j1}^{+(N)}\non\\ && {\cal T}_{1i}^{+(N)}=e^{i\mu m}  \sum_{j=i}^{n-1} T_{1j}^{+(N)}\ \hat T_{ji}^{+(N)} -i T_{11}^{+(N)}\ \hat T_{ni}^{+(N)},\non\\ && {\cal T}_{i1}^{+(N)}=e^{i\mu m}  \sum_{j=i}^{n-1} T_{ij}^{+(N)}\ \hat  T_{j1}^{+(N)} -i T_{in}^{+(N)}\ \hat  T_{11}^{+(N)}, ~~~i \in \{2, \ldots ,n \} \non\\ && {\cal T}_{kl}^{+(N)}=e^{i\mu m}\sum_{j=max(k,l)}^{n-1} T_{kj}^{+(N)}\ \hat T_{jl}^{+(N)}, ~~~k,l \in \{ 2, \ldots, n-1\}, \label{trep2} \ee  where $T_{ij}^{+(N)}$, $~\hat T_{ij}^{+(N)}$ are the entries of $T(\lambda \to \infty)$, $\hat T(\lambda \to \infty)$ (\ref{th}), i.e. tensor product realizations of $U_{q}(gl_{n})$ ({\it non}--affine). The charge associated to the affine generators is omitted here for brevity, but it is presented in \cite{doikou2}. The non--local charges (\ref{trep2}) form the  `boundary quantum algebra' satisfying exchange relations entailed from (\ref{re}) as $\lambda_{i} \to \infty$ \cite{doikou2}.
Boundary non-local charges were also identified in \cite{doikou3} in the isotropic case $q=1$ for two distinct types of boundary conditions, corresponding to boundary or twisted Yangians.

Our main aim now is to derive the conserved quantities commuting with the transfer matrix \cite{doikou2, doikou1}. Recall that we focus here on the case where the left boundary is trivial i.e. $K^{(l)} =I$.\\
\\
{\bf Proposition 2.} {\it 
The boundary charges (\ref{trep2}) in the fundamental representation commute with the  generators of ${\cal B}_{N}$ given in the representation (\ref{sol3})--(\ref{solb}) i.e. \be  \Big [\rho({\cal U}_{l}),\ \pi_{0}^{\otimes N}({\cal T}^{+(N)}_{ij}) \Big ] =0, ~~~l \in \{ 0, \ldots ,N-1 \}. \label{comte2} \ee} {\it Proof:} Recall that all the boundary charges are expressed in terms of the $U_{q}(gl_{n})$ generators it immediately follows that (\ref{comte2}) is valid for all $l \in {1, \ldots N-1}$.
It may be also proved by inspection, and given the tensor product form of the non-local charges, that  (\ref{comte2}) is valid for  $l=0$ as well.\\
\\
{\bf Corollary:} {\it The open spin chain Hamiltonian (\ref{ht}) commutes with the fundamental representation of the boundary charges (\ref{trep2}) \be \Big [{\cal H},\ \pi_{0}^{\otimes N}({\cal T}^{+(N)}_{ij}) \Big
] =0. \label{comh} \ee} This is evident from the form of ${\cal H}$ (\ref{ht}), which is expressed solely in terms of the representation $\rho({\cal U}_{l})$ of ${\cal B}_{N}$.

Also a general statement on the symmetry of the open transfer matrix can be made:\\
\\
{\bf Proposition 3.} {\it The open transfer matrix (\ref{transfer0}) commutes with all the non-local charges ${\cal T}^{+(N)}_{ij}$ (\ref{trep2}), i.e. \be  \Big [t(\lambda),\ {\cal T}^{+(N)}_{ij} \Big ] =0. \label{symm} \ee}
{\it Proof:}  The proof relies primarily on the existence of the generalized intertwining relations (\ref{it}) \cite{doikou2, doikou3, doikou1}. We shall use as a paradigm the $U_{q}(\widehat{sl_{2}})$ case \cite{doikou1}, although the proof may be generalized in a straightforward manner for the $U_{q}(\widehat{gl_{n}})$ case \cite{doikou2}. In the  $U_{q}(\widehat{sl_{2}})$ case there exist only one non-trivial boundary charge (non--affine) \cite{doikou1} i.e. \be {\cal T}_{11}^{+(N)}= q^{-{1\over 2}}K^{(N)}E^{(N)}+q^{{1\over 2}}K^{(N)}F^{(N)}+x_{1} (K^{(N)})^{2}-x_{1} I, \label{Q1} \ee where $K^{(N)}$, $~E^{(N)}$, $~F^{(N)}$ are the the $N$ coproducts of the $U_{q}(sl_{2})$ generators \cite{jimbo}. Let \be {\cal T}(\lambda) = \left(
\begin{array}{cc}
{\cal A}_{1} &  {\cal B}\\
{\cal C} &  {\cal A}_{2}   \\
\end{array} \right)  \label{tr2}
\ee then using (\ref{it}) as $\lambda' \to \infty$, and (\ref{tr2}) the following important algebraic relations are entailed \be \Big [{\cal T}_{11}^{+(N)},\ {\cal A}_{1} \Big ]= 
e^{-i\mu} ({\cal B}-{\cal C}),~~~~\Big [ {\cal T}_{11}^{+(N)},\ {\cal A}_{2} \Big ]= -e^{i\mu}({\cal B}-{\cal C}) \label{com0} \ee \be \Big [ {\cal T}_{11}^{+(N)},\ {\cal C} \Big ]_{q^{-1}} = 
{\cal A}_{2}-{\cal A}_{1} +x_{1}(q- q^{-1}){\cal C} \ee \be
\Big [{\cal T}_{11}^{+(N)},\ {\cal B} \Big ]_{q} =
{\cal A}_{1}- {\cal A}_{2} +x_{1}(q^{-1}- q){\cal B},  \label{com2} \ee where we define $[X,\ Y]_{q} =qX Y-q^{-1}Y X$, and $x_{1}$ is a constant depending on the boundary parameters $m,\ \zeta$. Recall that  $K^{(l)}(\lambda) =I$ and $M =diag(q,\ q^{-1})$, then the transfer matrix can be written as: \be t(\lambda) = e^{ i \mu} {\cal A}_{1} + e^{- i\mu} {\cal A}_{2}. \label{tt2} \ee Finally, by virtue of (\ref{tt2}) and recalling the exchange relations (\ref{com0})--(\ref{com2}) it follows that:
\be \Big [t(\lambda),\ {\cal T}_{11}^{+(N)} \Big ]= 0 \label{co2} \ee and this concludes our proof.\\
\\
\textbf{Acknowledgements:} This work is supported by the TMR Network `EUCLID', contract number HPRN-CT-2002-00325, and CNRS.


\begin{thebibliography}{99}

\bibitem{baxter} R.J. Baxter, {\it Exactly solved models in statistical mechanics} (Academic Press, 1982).

\bibitem{cherednik} I.V. Cherednik, Theor. Math. Phys. {\bf 61} (1984) 977.

\bibitem{jimbo} M. Jimbo, Lett. Math. Phys. {\bf 11} (1986) 247.

\bibitem{levymartin} D. Levy and P.P. Martin, J. Phys. {\bf A27} (1994) L521.

\bibitem{doikoumartin} A. Doikou and P.P. Martin, J. Phys. {\bf A36} (2003) 2203.

\bibitem{doikou2} A. Doikou, Nucl. Phys. {\bf B725} (2005) 439.

\bibitem{sklyanin} E.K. Sklyanin, J. Phys. {\bf A21} (1988) 2375.

\bibitem{dema} G. Delius and N. Mackay, Commun. Math. Phys. {\bf 233} (2003) 173.

\bibitem{doikou3} A. Doikou, J. Math. Phys.  {\bf 46}, 053504 (2005).

\bibitem{doikou1} A. Doikou, math-ph/0402067.



\end{thebibliography}
\end {document}